\documentclass[twocolumn,aps,prb,superscriptaddress,floatfix]{revtex4-1}
\usepackage{graphicx}
\usepackage{polski}
\usepackage[utf8]{inputenc}
\usepackage{dcolumn}
\usepackage[T1]{fontenc}
\usepackage{amsmath}
\usepackage{sidecap}

\begin{document}
\title{High frequency voltage-induced ferromagnetic resonance in magnetic tunnel junctions}

\author{Witold Skowro\'{n}ski}
 \email{skowron@agh.edu.pl}
\author{Stanisław Łazarski}
\author{Jakub Mojsiejuk}
\author{Jakub Chęciński}
\author{Marek Frankowski}
\affiliation{AGH University of Science and Technology, Department of Electronics, Al. Mickiewicza 30, 30-059 Krak\'{o}w, Poland}
\author{Takayuki Nozaki}
\author{Kay Yakushiji}
\author{Shinji Yuasa}
\affiliation{National Institute of Advanced Industrial Science
and Technology, Spintronics Research Center, Tsukuba, Ibaraki 305-8568, Japan}

\begin{abstract}
Voltage-induced ferromagnetic resonance (V-FMR) in magnetic tunnel junctions (MTJs) with a W buffer is investigated. Perpendicular magnetic anisotropy (PMA) energy is controlled by both thickness of a CoFeB free layer deposited directly on the W buffer and a post-annealing process at different temperatures. The PMA energy as well as the magnetization damping are determined by analysing field-dependent FMR signals in different field geometries. An optimized MTJ structure enabled excitation of V-FMR at frequencies exceeding 30 GHz. The macrospin modelling is used to analyse the field- and angular-dependence of the V-FMR signal and to support experimental magnetization damping extraction.
\end{abstract}

\maketitle

\section{Introduction}
\label{sec:intro}
Magnetic multilayer structures are commonly utilized as key elements of magnetic field sensors \cite{tumanski_thin_2001} and magnetic random access memories \cite{bhatti_spintronics_2017}. A stack structure similar to the one used in existing applications and related physical mechanisms, such as magnetoresistance effect and spin-transfer-torque (STT), can be used for future microwave solutions, for example spin torque oscillators \cite{chen_sto_review_2015} or microwave detectors \cite{skowronski_microwave_2016}. One of the main drawbacks of the existing technologies is a relatively high current used in the STT-based devices. In order to reduce the power consumption of such devices, a few alternative solutions have been proposed, including spin-orbit torque \cite{liu_spin-torque_2012, miron_current-driven_2010} and electric-field controlled magnetism \cite{weisheit_electric_2007, Miwa_perpendicular_2018}. The latter typically requires structures with relatively thick insulators for application of a significant electric-field at the insulator/ferromagnet interface.

Recently, a few different mechanisms have been proposed to maximize the effect of an electric field on magnetic properties of materials, including diluted magnetic semiconductors \cite{chiba_magnetization_2008}, nitride semiconductors \cite{sztenkiel_stretching_2016}, charge migration in multilayers \cite{bauer_electric_2012} and voltage controlled magnetic anisotropy (VCMA) in metallic thin films \cite{maruyama_large_2009}. It has been already presented that, by utilizing the VCMA effect, driving the magnetization into a precession at several GHz is possible \cite{zhu_voltage-induced_2012}, which is promising for high-frequency devices. 

In this work we present studies on CoFeB/MgO/ CoFeB-based MTJs deposited on a W buffer \cite{skowronski_underlayer_2015, chang_electric_2017}. Using a relatively thin CoFeB bottom free layer and different annealing temperatures, high perpendicular magnetic anisotropy (PMA) energies of up to 1.5 MJ/m$^3$ are achieved, which, in turn, enables voltage-induced ferromagnetic resonance (V-FMR) excitation at frequencies exceeding 30 GHz. In addition, the V-FMR measurements in combination with vector network analyser ferromagnetic resonance (VNA-FMR) investigation \cite{bilzer_study_2006, glowinski_coplanar_2014} were used to determine the magnetization damping in the discussed multilayers.

\section{Experiment}
\label{sec:experiment}
Multilayers of the following structure: W(5)/ CoFeB($t_\mathrm{CoFeB}$)/ MgO(1.8)/ CoFeB(5)/ Ta(5)/ Ru(5)/ Pt(2) (thickness in nm) were deposited using magnetron sputtering in a similar conditions to Ref. \cite{skowronski_perpendicular_2015}. Two thicknesses of the free layer, i.e., $t_\mathrm{CoFeB}$ = 0.9 and 1.2 nm, were investigated. After the deposition process, magnetic properties were determined with VNA-FMR in 10 $\times$ 10 mm samples by analysing the frequency vs. perpendicular magnetic field dependence using the Kittel relation $f$ = $\gamma$/(2$\pi$)$\sqrt{BH}$, where $H$ is the external magnetic field and $B$ is the magnetic field induction. Measurements were repeated after subsequent sample annealing in high-vacuum furnace at $T_\mathrm{A}$ = 200, 250, 300 and 350$^\circ$C. Independently, the same multilayers were fabricated into pillars of 2 $\times$ 4 $\mu$m$^2$ using electron-beam lithography and ion-beam milling for transport measurements. The transport properties were measured in a probe station that enabled sample rotation at azimuth ($\theta$) and polar ($\phi$) angle with respect to the sample plane, in a magnetic field of up to 1 T with a broadband electrical contact. Both quasi-static (resistance vs. magnetic field) and dynamic measurements (DC mixing voltage vs. magnetic field) were performed using a two-point method.

\begin{table}[!t]
\caption{\label{tab:table1}%
PMA energies of samples with different $t_\mathrm{CoFeB}$ after annealing at increasing temperatures. Values of $K_\mathrm{PMA}$ in bold indicate samples with perpendicular effective anisotropy.}
\begin{ruledtabular}
\begin{tabular}{ l c c}
  &
 $t_\mathrm{CoFeB}$ = 0.9 nm &
 $t_\mathrm{CoFeB}$ = 1.2 nm
 \\
  $T_\mathrm{A}$ ($^\circ$C) &
 $K_\mathrm{PMA}$ (MJ/m$^3$)&
 $K_\mathrm{PMA}$ (MJ/m$^3$)
 \\ 
 \colrule
as dep. & 0.5 & 0.4 \\
200 & 0.7 & 0.6 \\
250 & \textbf{1.04} & 0.8 \\
300 & \textbf{1.19} & 0.95 \\
350 & \textbf{1.51} & \textbf{1.12} \\
\colrule
\end{tabular}
\end{ruledtabular}
\end{table}

\section{Results and discussion}
\label{sec:results}
\begin{figure}
\centering
\includegraphics[width=\columnwidth]{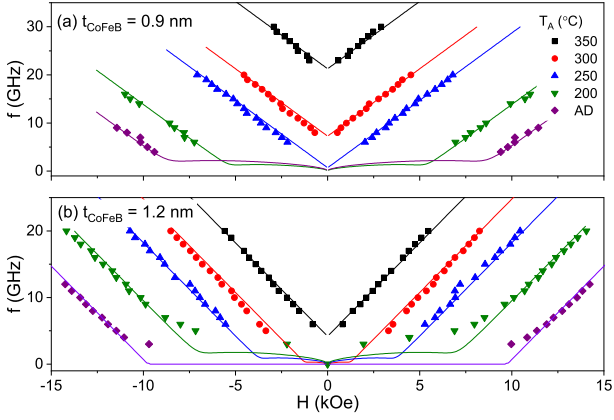}
\caption{$f$ vs. $H$ dependence of samples with $t_\mathrm{CoFeB}$ = 0.9 (a) and $t_\mathrm{CoFeB}$ = 1.2 nm (b) measured after annealing at different temperatures using VNA-FMR  - points. Lines represents best fits to the model based on the Kittel formula, resulting in $K_\mathrm{PMA}$ summarized in Table \ref{tab:table1}.}
\label{fig:fig1}
\end{figure}

First, we focus on the wafer-level investigation using VNA-FMR. Resonance frequency ($f$) vs. perpendicular magnetic field curves of samples with different thicknesses of the free layer ($t_\mathrm{CoFeB}$ = 0.9 and 1.2 nm) are presented in Fig. \ref{fig:fig1}. Experimental points were modelled using the Kittel relation with a CoFeB saturation magnetization value of $\mu_0M_S$ = 1.6 T \cite{skowronski_underlayer_2015}. The PMA energies ($K_\mathrm{PMA}$) resulting from fits to the model are gathered in Table \ref{tab:table1}. Assuming an infinite-plane sample configuration, the demagnetization energy density $K_\mathrm{D}$ = $\mu_0M_\mathrm{S}^2$/2 = 1.02 MJ/m$^3$, resulting in an effective perpendicular anisotropy for the sample with $t_\mathrm{CoFeB}$ = 0.9 nm for $T_\mathrm{A}$ $\geq$ 250$^\circ$C and for sample with $t_\mathrm{CoFeB}$ = 1.2 nm for $T_\mathrm{A}$ $>$ 300$^\circ$C. The PMA energy of the sample with $t_\mathrm{CoFeB}$ = 0.9 nm annealed at $T_\mathrm{A}$ = 350$^\circ$C evaluated from the TMR dependence on the magnetic field applied in the sample plane \cite{skowronski_perpendicular_2015} is K = 1.47 MJ/m$^3$, which is in good agreement with the values obtained from VNA-FMR.

\begin{figure}[t!]
\centering
\includegraphics[width=\columnwidth]{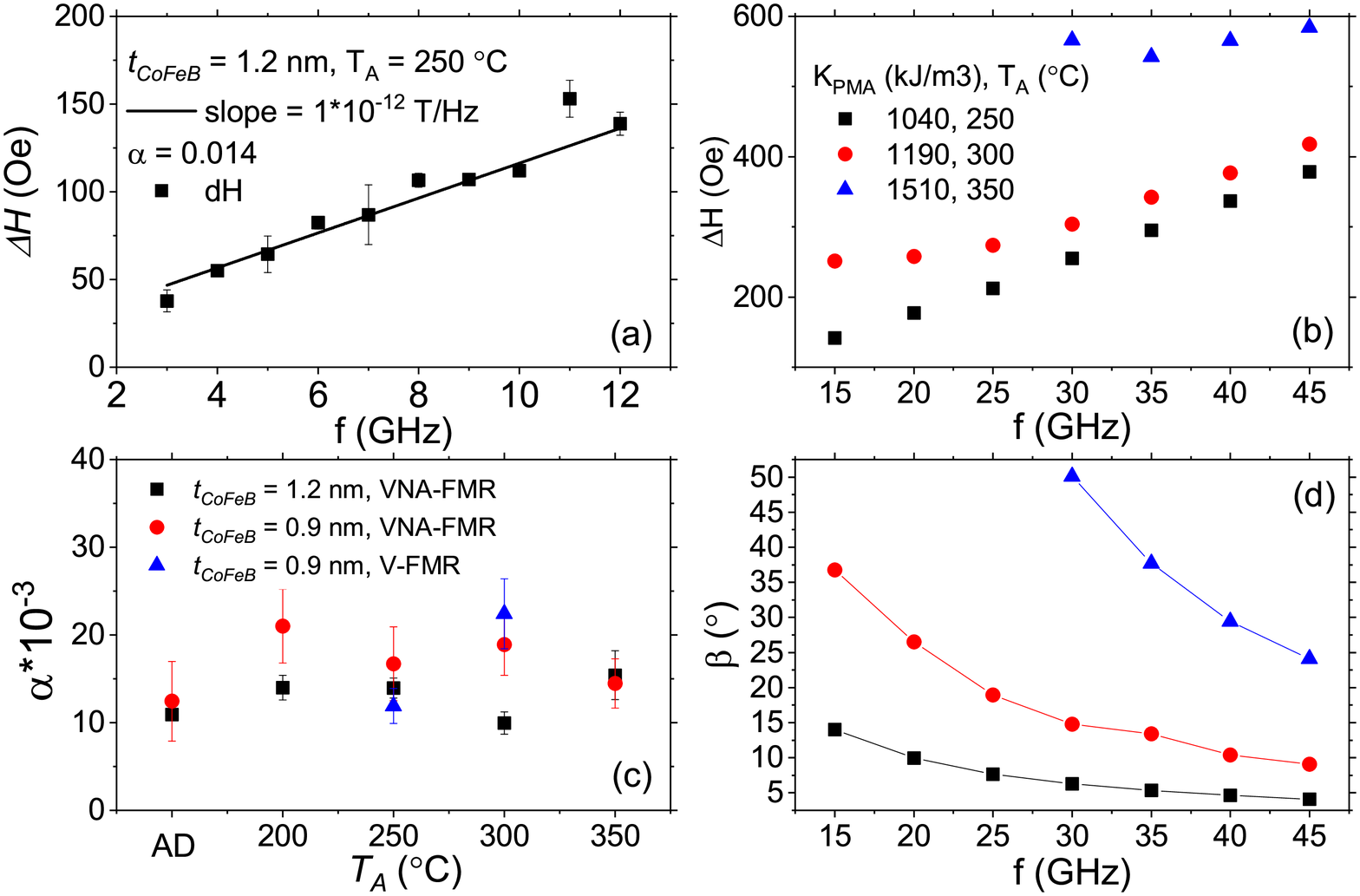}
\caption{(a) Example of $\Delta H$ vs. $f$ dependence for the  sample with $t_\mathrm{CoFeB}$ = 1.2 nm annealed at 250 $^\circ$C measured using VNA-FMR. Fitting a line resulted in slope of 10$^{-12}$ T/Hz and $\alpha$ = 0.014. The linewidth vs. $f$ dependence, presented in (b), indicates a non-linear behaviour for small $f$ (and thus small $H$). (c) presents summarized $\alpha$ for all samples obtained using both VNA-FMR and V-FMR. The origin of a non-linear $\Delta H$ vs. $f$ is explained in (d), where $\beta$ is the angle between the applied magnetic field vector and the magnetization vector - a strong non-collinearity is found in simulations for small $f$ and high $K_\mathrm{PMA}$.}
\label{fig:fig2}
\end{figure}

From the same type of measurements, the free layer magnetization damping ($\alpha$) was calculated from the linewidth ($\Delta H$) vs. $f$ slope according to the following Eq.;
\begin{equation}
\label{eq:alpha}
\Delta H = \frac{4 \pi \alpha}{\gamma}f + \Delta H_0
\end{equation}
where $\gamma$ = 1.76$\times$10$^{11}$ Hz/T is the gyromagnetic ratio and $\Delta H_0$ is the inhomogeneous broadening \cite{shaw_resolving_2014}. An example of the $\Delta H$ vs. $f$ dependence for the sample with $t_\mathrm{CoFeB}$ = 1.2 nm annealed at $T_\mathrm{A}$ = 300$^\circ$C is presented in Fig. \ref{fig:fig2}(a). The calculated $\alpha$ for all samples is presented in Fig. \ref{fig:fig2}(c). The values obtained for the sample with $t_\mathrm{CoFeB}$ = 1.2 nm agree well with the W/CoFeB sample of similar thickness reported in Ref. \cite{couet_impact_2017}. One can also note a tendency of $\alpha$ to increase with decreasing thickness of CoFeB \cite{swindells_spin_2019}.

\begin{figure}[!ht]
\centering
\includegraphics[width=\columnwidth]{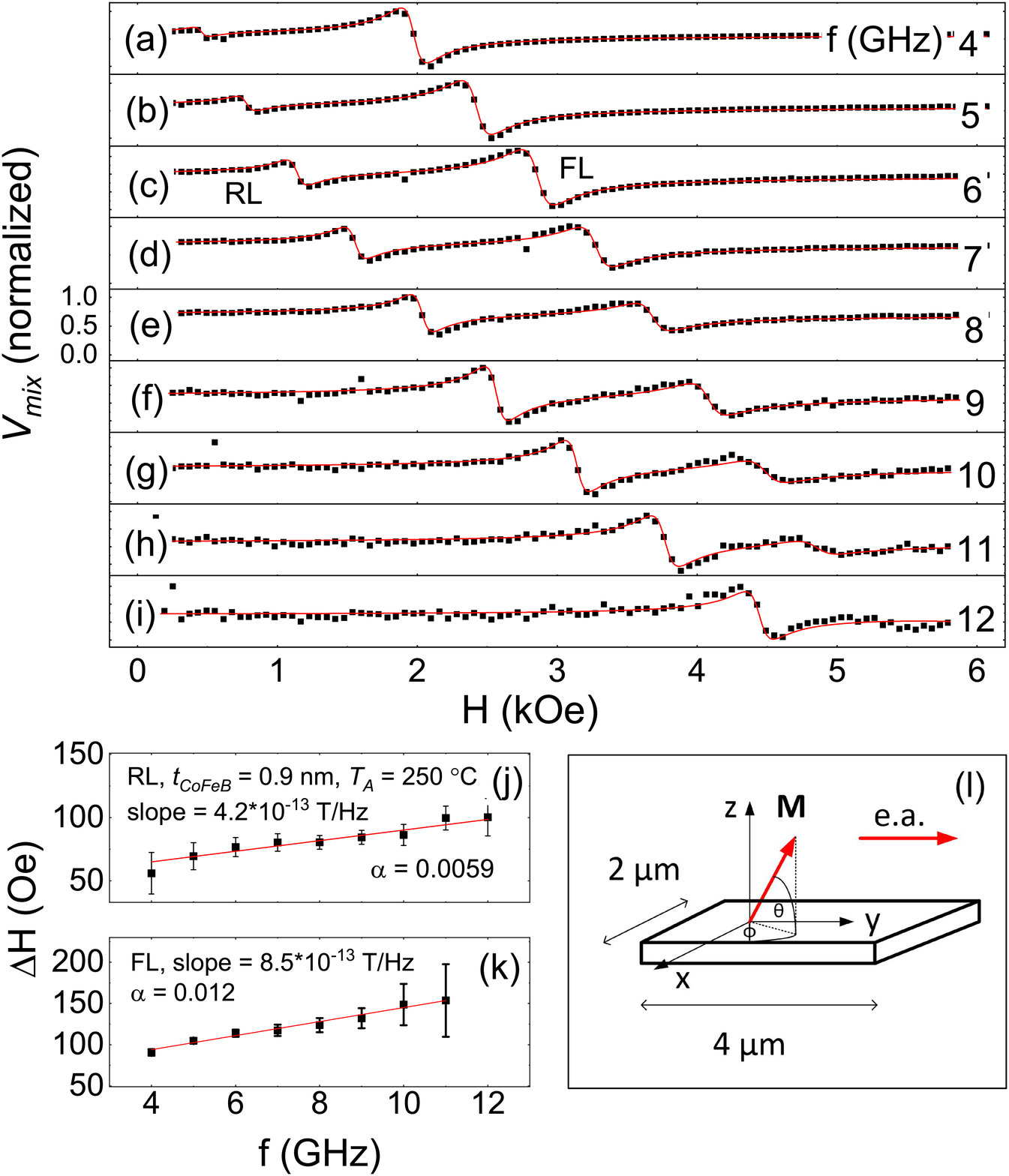}
\caption{(a-i) $V_\mathrm{mix}$ vs. $H$ applied at $\theta$ = 60$^\circ$ and $\phi$ = 90$^\circ$ dependence measured with an excitation frequency $f$ between 4 and 12 GHz, for sample with $t_\mathrm{CoFeB}$ = 0.9 nm annealed at 250 $^\circ$C. Lines represent best fits to Eq. \ref{eq:vmix}. (j) $\Delta H_1$ vs. $f$ dependence for the reference layer (RL), fitting a line resulted in slope of 4.2$\times$10$^{-13}$ T/Hz and $\alpha$ = 0.006,  (k) $\Delta H_2$ vs. $f$ dependence of the free layer (FL), fitting a line resulted in slope of 8.5$\times$10$^{-13}$ T/Hz and $\alpha$ = 0.012, (l) presents a sketch of the experimental geometry}
\label{fig:fig3}
\end{figure}

Next, we move on to the transport measurement on the fabricated sample. The crystallization process of CoFeB (initiated from the MgO tunnel barrier) starts after annealing at 250$^\circ$C \cite{yuasa_giant_2007}. Therefore, we begin the analysis of the V-FMR signal for the samples annealed at this temperatures. The tunnelling magnetoresistance ratio of the fabricated MTJ reaches 40\% (measured between an orthogonal and  parallel CoFeB magnetization orientations) and the resistance-area product is around $RA$ = 40 kOhm*$\mu$m$^2$. 
Figure \ref{fig:fig3}(a-i) presents DC voltage (a result of a small AC current mixing with resistance change originating from the VCMA) $V_\mathrm{mix}$ as a function of the magnetic field applied at an angle $\theta$ = 60$^\circ$ from the sample plane. The resonance peak at higher (smaller) field originates from the free (reference) layer. This dependence was modelled using a sum of two symmetric and asymmetric Lorentz curves:   

\begin{equation}
\begin{split}
V_{\text{mix}}=V_{S1}\frac{\Delta H_1^2}{(H-H_{r1})^2+\Delta H_1^2} \\ + V_{A1}\frac{(H-H_{r1})\Delta H_1}{(H-H_{r1})^2+\Delta H_1^2} \\+ V_{S2}\frac{\Delta H_2^2}{(H-H_{r2})^2+\Delta H_2^2} \\ + V_{A2}\frac{(H-H_{r2})\Delta H_2}{(H-H_{r2})^2+\Delta H_2^2} \\
\end{split}
\label{eq:vmix}
\end{equation}
where $H_\mathrm{r1}$ ($H_\mathrm{r2}$) is the resonance field of the free (reference) layer at a given $f$, $V_\mathrm{S1}$ and $V_\mathrm{A1}$ ($V_\mathrm{S2}$ and $V_\mathrm{A2}$) are the amplitudes of the symmetric and antisymmetric Lorentz functions of the free (reference) layer and $\Delta H_1$ ($\Delta H_2$) is the linewidth of the free (reference) layer. Fitting the $\Delta H_1$ and $\Delta H_2$ vs. $f$ dependence enabled us to calculate the damping of the free and the reference layers independently - Fig. \ref{fig:fig3}(j-k). The results agree well with $\alpha$ obtained using the VNA-FMR method and are included in Fig. \ref{fig:fig2}(c).

\begin{figure}[!ht]
\centering
\includegraphics[width=\columnwidth]{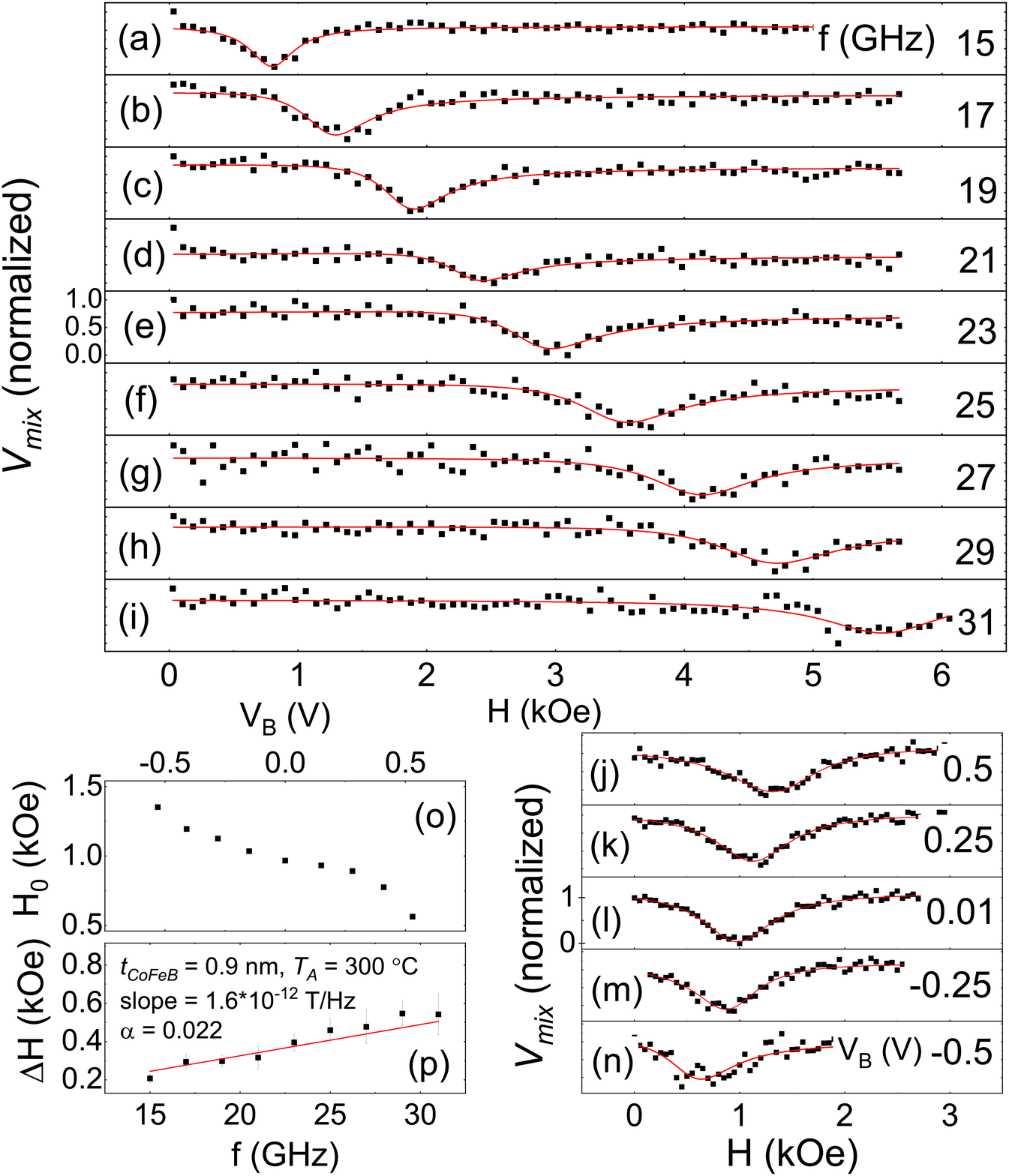}
\caption{(a-i) $V_\mathrm{mix}$ vs. $H$ applied at $\theta$ = 85$^\circ$ and $\phi$ = 0$^\circ$ with different $f$ of the excitation signal of sample with $t_\mathrm{CoFeB}$ = 0.9 nm annealed at 300 $^\circ$C. Lines represent best fits to Eq. \ref{eq:vmix}, (j-n) shows $V_\mathrm{mix}$ vs. $H$ dependence for $f$ = 17 GHz under different bias voltage condition, (o) shows the dependence of the resonance frequency on $V_\mathrm{B}$, (p) $\Delta H_1$ vs. $f$ dependence, fitting a line resulted in slope of 1.6$\times$10$^{-12}$ T/Hz and $\alpha$ = 0.024}
\label{fig:fig4}
\end{figure}

Within the limit of the available magnetic field of up to 1 T, for the MTJ annealed at 250$^\circ$C the maximum resonance signal was measured for around $f$ = 20 GHz, depending on the magnetic field configuration \cite{bonetti_spin_2009}. In order to increase the resonance frequency, the fabricated sample was further annealed at 300$^\circ$C, which enhanced the PMA energy. The V-FMR signal measured at a nearly perpendicular field ($\theta$ = 85$^\circ$) is presented in Fig. \ref{fig:fig4}(a-i). A much better signal-to-noise ratio was obtained for an azimuth angle $\phi$ = 0$^\circ$, resulting in a symmetric lineshape, contrary to the signal presented in Fig. \ref{fig:fig3}(a-i), where $\phi$ = 90$^\circ$ resulted in an asymmetric signal. In the general case, the lineshape depends on the angle of the free layer magnetization vector projection on the sample plane and the reference layer magnetization, which, in our case is defined along the $y$ axis (Fig. \ref{fig:fig3}(l)) \cite{nozaki_electric-field-induced_2012}. We also note that, contrary to the previous studies of V-FMR in MTJs \cite{shiota_high_2014, kanai_electric_2014}, the lineshape is only little affected by the bias voltage ($V_\mathrm{B}$), which is due to strong PMA in our devices - Fig. \ref{fig:fig4}(j-n). Therefore, magnetic anisotropy changes induced by static bias voltage are weak comparing to high PMA energy.

To understand the angular dependence of the V-FMR and to confirm the magnetization damping analysis, the macrospin simulations were performed. We used Landau-Lifshitz-Gilbert equation, where voltage excitation was modelled as sinusoidal changes of $K_\mathrm{PMA}$. This, in turn, contributes an alternating term to the effective field, which determines magnetization dynamics. Afterwards, MTJ resistance is calculated and multiplied by the assumed leakage current associated with the applied voltage, resulting in $V_\mathrm{mix}$, in the same way as in the experimental procedure. We used a similar approach previously in V-FMR spin diode effect modelled using micromagnetic simulations \cite{frankowski_perpendicular_2017}. 
 
The simulated dependence of the linewidth vs. $f$ for different $K_\mathrm{PMA}$ is depicted in Fig. \ref{fig:fig2}(b). A strong deviation from the linear dependence is observed for high $K_\mathrm{PMA}$, which is a result of a significant non-collinear direction of the magnetic field and magnetization in this case presented in Fig. \ref{fig:fig2}(d). 
At the same time, magnetic fields sufficient to saturate the sample would increase the effective field to the level where the VCMA excitation would no longer be strong enough to induce significant magnetization dynamics \cite{frankowski_perpendicular_2017}. Therefore, we limit the V-FMR investigation to samples annealed at 250 and 300$^\circ$C. 




\section{Summary}
\label{sec:summary}
In conclusion, ferromagnetic resonance in W/CoFeB/MgO/CoFeB was investigated by means of wafer-level vector network analyser FMR and voltage-induced FMR in patterned devices. Both the CoFeB thickness and the thermal treatment influence magnetic anisotropy, which reaches a value of 1.5 MJ/m$^3$, which is well above demagnetizing energy. Resonance signals from both the reference and the free layer are analysed, allowing for magnetization damping determination. For thin CoFeB free layers the damping between 0.01 and 0.02 was measured, independent on annealing conditions, which is approximately double of the thicker reference layer damping. V-FMR in the MTJ annealed at 300 $^\circ$C is measured up to a high value of $f$ = 31 GHz. 

\section*{Acknowledgments}
We would like to thank prof. Tomasz Stobiecki, prof. Sławomir Gruszczyński and dr. Sławomir Ziętek for a fruitful discussions and technical assistance in the measurements. This work was supported by the National Science Centre, Poland, grant No. LIDER/467/L-6/14/NCBR/2015 by the Polish National Centre for Research and Development. M.F. acknowledges grant Preludium UMO-2015/17/N/ST3/02276 from National Science Center, Poland. Microfabrication was performed at Academic Centre for Materials and Nanotechnology of AGH University. Numerical calculations were supported by PL-GRID infrastructure.

\bibliographystyle{apl}
\bibliography{Skowronski_library}
\end{document}